\documentclass[12pt,english]{article}
\usepackage{amsmath}
\usepackage{amssymb}
\usepackage{mathrsfs}
\usepackage{graphicx}
\usepackage{fullpage}

\makeatletter

\catcode`@=12 \topmargin 0.5in \evensidemargin 0.0in
\usepackage{babel}
\makeatother
\newcommand {\be}{\begin{equation}}
\newcommand {\ee}{\end{equation}}
\newcommand {\ba}{\begin{eqnarray}}
\newcommand {\ea}{\end{eqnarray}}

\begin{document}
\thispagestyle{empty}
\begin{flushright}
IPM/P-2010/001\\
\today
\end{flushright}

\mbox{} \vspace{0.75in}

\begin{center}\textbf{\large  Extracting  the CP-violating phases of trilinear
R-parity violating couplings from
$\mu \to eee$}\\

 \vspace{0.5in} \textbf{ $\textrm{Yasaman Farzan}$\footnote{yasaman@theory.ipm.ac.ir} and \textrm{Saereh Najjari} $^\S$}
\\
 \vspace{0.2in}\textsl{School of physics, Institute for Research in Fundamental Sciences (IPM)\\P.O. Box 19395-5531, Tehran, IRAN}\\
 \vspace{0.2in}\textsl{${}^\S$ Physics Department, Tafresh University, Tafresh, IRAN}\\

 \vspace{.75in}\end{center}

\baselineskip 20pt
\begin{abstract}
It has recently  been  shown that by measuring the transverse
polarizations of the final particles in $\mu^+ \to e^+ e^- e^+$,
it is possible to extract  information on the phases of the
effective couplings leading to this decay. We examine this
possibility within the context of $R$-parity violating Minimal
Supersymmetric Standard Model (MSSM) in which the $\mu^+ \to e^+
e^-e^+$ process can take place at a tree level. We demonstrate how
a combined analysis of the angular distribution of the emitted
electrons and their transverse polarization can determine the
CP-violating phases of the trilinear R-parity violating Yukawa
couplings.

\end{abstract}

\section{Introduction}
If  the neutrino masses are the only sources of Lepton Flavor
Violation (LFV), the rates of
 the Lepton
Flavor Violating  processes such as $\mu \to e\gamma$, $\mu N \to
e N$ and $\mu \to eee$ will be too small to be detectable
 in the foreseeable future. Any
positive signal for such LFV processes will be an indication for new
physics beyond the  Standard Model (SM). Strong experimental bounds exist on the rate
of these processes \cite{pdg} and rich literature has been developed
on the constraints on new physics from these bounds. Currently, the
MEG experiment at PSI is searching for $\mu \to e \gamma$ and will
be able to detect a signal even if Br$(\mu \to e \gamma)$ is as
small as $O(10^{-13})$.

 We can divide the beyond SM scenarios into two
classes: (1)  Models, such as R-parity conserving Minimal
Supersymmetric Standard Model (MSSM), within which only even
numbers of new particles can appear in each vertex; (2) Models,
such as R-parity violating MSSM, within which the number of new
particles in each vertex can be both even or odd. In case of the
first class of models, both $\mu \to e \gamma$ and $\mu \to eee$
can take place only at the loop level. Within these models Br($\mu
\to eee$), being a three body decay, is typically smaller than
Br($\mu \to e \gamma$) so the latter gives stronger bounds on the
LFV parameters ({\it see}, however \cite{deGouvea:2000cf}).
However, within the second class of the models, $\mu \to eee$ can
take place at tree level and its rate can therefore exceed that of
$\mu \to e \gamma$ which is possible only at a loop level. Within
the R-parity violating MSSM, it is possible that while Br($\mu \to
e \gamma$) is too small to be probed even by MEG, Br($\mu \to
eee$) is relatively large and close to its present bound
\cite{deGouvea:2000cf}. A mild improvement on $\mu \to eee$ can
probe R-parity violating MSSM with mass scale well above 10 TeV
which is beyond the reach of the LHC.

Recently it has been shown that by measuring the transverse
polarizations of the final particles in $\mu \to e \gamma$,
$\mu-e$ conversion on nuclei and $\mu \to eee$, one can measure
the CP-violating phases of the general effective Lagrangian
leading to these processes {\cite{eee,Ayazi:2008gk,tracing}. The
possibility of deriving information on the CP-violating phases of
the R-parity conserving MSSM from $\mu \to e \gamma$ and the $\mu
-e$ conversion has been explored in \cite{Ayazi:2008gk}. In the
present {\it letter}, we are going to examine the possibility of
deriving the CP-violating phases of the trilinear R-parity
violating couplings from $\mu \to eee$. These CP-violating phases
are important as they can be responsible for the creation of the
baryon asymmetry in the universe \cite{RprimeRparity}.

The paper is organized as follows. In section~2, we review the
general effective Lagrangian that can lead to $\mu \to eee$ and
$\mu \to e\gamma$. We outline the information that can be derived
on the effective couplings without the measurement of the spin of
the final particles. In section~3, we review the contributions
that the effective couplings receive from the $R$-parity and
lepton flavor violating  trilinear Yukawa couplings. We then
briefly review bounds on these couplings. In section~4, we
introduce the P-odd asymmetry, $\mathcal{A}$, and the transverse
polarization. In section~5, we briefly discuss the feasibility of
measuring the transverse polarization of the final particles. In
section~6, we discuss the results that can be derived by combined
analysis of $\mathcal{A}$ and the transverse polarization. Results
are summarized in section~7.

\section{ Effective Lagrangian in a general framework}
The effective Lagrangian leading to $\mu \to eee$ can in general
be written as \cite{review} \ba \label{effectiveLag}{\cal L}_{\rm eff}&=& B_1
(\bar{\mu}_Le_R)( \bar{e}_R e_L)+B_2( \bar{\mu}_R e_L
)(\bar{e}_Le_R) + \ \ \ \cr &\ &
 C_1
(\bar{\mu}_Le_R)( \bar{e}_L e_R)+C_2 (\bar{\mu}_Re_L)( \bar{e}_R
e_L)+ \cr ~ &\ & G_1 (\bar{\mu}_R\gamma^\nu e_R)(\bar{e}_R\gamma_\nu
e_R)+G_2 (\bar{\mu}_L\gamma^\nu e_L)(\bar{e}_L\gamma_\nu e_L) \cr &
\ &
-A_R~(\bar{\mu}_L[\gamma_\mu,\gamma_\nu]~\frac{q^\nu}{q^2}e_R)~(\bar{e}\gamma^\mu
e)-A_L~(\bar{\mu}_R[\gamma_\mu,\gamma_\nu]~\frac{q^\nu}{q^2}e_L)~(\bar{e}\gamma^\mu
e)+{\rm H.c.}
\ea Notice that by using the identities
$(\sigma^\mu)_{\alpha \beta}(\sigma_\mu)_{\gamma \delta}\equiv 2
\epsilon_{\alpha \gamma} \epsilon_{\beta \delta}$ and $
(\bar{\sigma}^\mu)_{\alpha \beta}=\epsilon_{\beta \delta}
(\sigma^\mu)_{\delta \gamma} \epsilon_{\gamma \alpha}$  (where
$\epsilon_{11}=\epsilon_{00}=0$ and
$\epsilon_{10}=-\epsilon_{01}=1$) and employing the fact that the
fermions anti-commute, we can rewrite the terms on the first line
of Eq.~(\ref{effectiveLag}) as
$$-{B_1 \over 2}(\bar{\mu}_L  \gamma^\nu e_L)(
\bar{e}_R  \gamma_\nu e_R) -{B_2 \over 2}(\bar{\mu}_R  \gamma^\nu
e_R)( \bar{e}_L \gamma_\nu e_L) \ .$$ This effective Lagrangian
leads to \cite{review} \ba Br(\mu \to eee)&=&{1 \over 32
G_F^2}\left[|B_1|^2+|B_2|^2+8(|G_1|^2+|G_2|^2)\right.\cr
&+&{|C_1|^2+|C_2|^2 \over 2}+32(4\log {m_\mu^2\over
m_e^2}-{11}){|A_R|^2+|A_L|^2 \over m_\mu^2} \cr  &-& \left.
64{\Re[A_LG_2^*+A_RG_1^*]\over
m_\mu}+32{\Re[A_LB_1^*+A_RB_2^*]\over m_\mu}\right]\
.\label{EFFlag}\ea By studying the energy distribution of the
final particles, more information can be derived on the effective
couplings. For example, let us consider the contributions  from
$A_L$ and $A_R$ which come from a virtual photon exchange. When
the invariant mass of an electron positron pair goes to zero, the
virtual exchanged photon goes on shell. As a result, the
corresponding Dalitz plot should have a peak at
$(P_{e^-}+P_{e^+})^2=0$ whose height is given by
$|A_L|^2+|A_R|^2$. The combinations that can be derived by
studying the energy distributions of the final particles are
$|G_1|^2+|G_2|^2+(|C_1|^2+|C_2|^2)/16$, $|B_1|^2+|B_2|^2$,
$|A_L|^2+|A_R|^2$, ${\rm Re}[A_LG_2^*+A_R G_1^*]$ and ${\rm
Re}[A_LB_1^*+A_RB_2^*]$ (see for example \cite{review} and
references therein). By studying the angular distributions of the
final particles relative to the polarization of the initial muon,
one can further derive $|G_1|^2-|G_2|^2+(-|C_1|^2+|C_2|^2)/16$,
$|B_1|^2-|B_2|^2$, $|A_L|^2-|A_R|^2$, ${\rm Re}[A_LG_2^*-A_R
G_1^*]$ and ${\rm Re}[A_LB_1^*-A_RB_2^*]$ as well as the CP-odd
quantities ${\rm Im}[A_LG_2^*+A_R G_1^*]$ and ${\rm
Im}[A_LB_1^*+A_RB_2^*]$. A combined analysis of angular and energy
distribution therefore yields the absolute values of all the
effective couplings, $B_1$, $B_2$, $A_L$ and $A_R$ as well as the
CP-violating phases $\arg[A_LA_R^* G_2^* G_1]$, $\arg[A_LG_2^*]$,
$\arg[A_LA_R^* B_1^* B_2]$ and  $\arg[A_LB_1^*]$. Notice however
that there is still some information in Eq.~(\ref{effectiveLag})
that cannot be derived by the methods described above. In
particular, the CP-violating phase arg$[B_1B_2^*]$ cannot be
derived by these methods.  Further information can be obtained by
studying the transverse polarization of $e^-$ in $\mu^+\to
e^+e^-e^+$ \cite{eee}.

Notice that the terms on the last line of Eq.~(\ref{effectiveLag})
come from the effective coupling of the photon which gives rise to
\be \label{muegamma} {\rm Br}(\mu \to e \gamma)=\frac{12 \pi }{
G_F^2 m_\mu^2\alpha} \left(|A_L|^2+|A_R|^2 \right) \ .\ee The
present bound on this LFV rare decay is very strong ${\rm Br}(\mu
\to e \gamma)<1.2\times 10^{-11}$ \cite{pdg} which implies
$|A_L|^2+|A_R|^2<3.3\times{10^{-27}}~{\rm GeV}^{-2}$. Thus, from
Eq.~(3), we observe that the contribution from $A_L$ and $A_R$ to
$\mu \to eee$ cannot be larger than $7 \times 10^{-14}$ so if
Br($\mu \to eee$) turns out to be close to its present bound, we
can safely neglect the contributions from $A_L$ and $A_R$.

\section{Effects of trilinear R-parity violating couplings}
Within the R-parity conserving MSSM, the  slepton mass matrix  as
well as the trilinear $A$-term involving the sleptons include LFV
sources which can induce  the effective Lagrangian in
Eq.~(\ref{effectiveLag}). By relaxing the R-parity conservation,
new sources of LFV emerge. In particular, the R-parity violating
Yukawa couplings in the superpotential \be W={\lambda_{ijk}\over
2} \hat{L}_i \hat{L}_j \hat{E}_k ~~~~~{\rm with}~~~~~
\lambda_{ijk}=-\lambda_{jik} \  \ee add nine new sources of LFV.
That is each nonzero element of $\lambda_{ijk}$ is a source of
LFV. Some of these couplings can contribute to $\mu \to eee$ at
tree level. Throughout this letter, we will focus on the effects
of $\lambda$ and set other LFV sources equal to zero for
simplicity. The nonzero elements of $\lambda_{ijk}$ contain nine
phases out of which three can be absorbed by rephasing
$\hat{L}_i$. In this basis, each of the  bilinear R-parity terms,
$\mu_i^\prime \hat{L}_i \hat{H}_u$  can be considered as a new
source for CP-violation. In this {\it letter}, we will investigate
the possibility of deriving the CP-violating phases of
$\lambda_{ijk}$ from the transverse polarization of the final
particles in $\mu \to eee$.

The contributions to the effective couplings from $\lambda_{ijk}$
have been calculated in \cite{deGouvea:2000cf} and the results are
as follows:
\begin{subequations} \label{Cs}
  \begin{align}
  A_L&=G_2=C_1=C_2=0,\\ \label{Cb}
B_{1}&=\frac{\lambda_{321}^{*}\lambda_{311}}{m^{2}_{\tilde{\nu}_{\tau}}},\\
\label{Cc}
B_{2}&=\frac{\lambda_{211}^{*}\lambda_{212}}{m^{2}_{\tilde{\nu}_{\mu}}}
+\frac{\lambda_{311}^{*}\lambda_{312}}{m^{2}_{\tilde{\nu}_{\tau}}} \\
&-\frac{\alpha}{12\pi} \left[\frac{\lambda_{321}^{*}\lambda_{311}}
{m^{2}_{\tilde{\nu}_{\tau}}}
(-\frac{8}{3}-2\log\frac{m^2_e}{m^{2}_{\tilde{\nu}_{\tau}}}
+\frac{m^2_{\tilde{\nu}_{\tau}}}{3m^2_{\tilde{e}_{R}}})\right.\nonumber\\
&~~~~~~~~+\frac{\lambda_{322}^{*}\lambda_{312}}{m^{2}_{\tilde{\nu}_{\tau}}}
(-\frac{8}{3}-2\log\frac{m^2_\mu}{m^{2}_{\tilde{\nu}_{\tau}}}
+\frac{m^2_{\tilde{\nu}_{\tau}}}{3m^2_{\tilde{\mu}_{R}}})
\nonumber\\ &~~~~~~~~+\left.
\frac{\lambda_{323}^{*}\lambda_{313}}{m^{2}_{\tilde{\nu}_{\tau}}}
(-\frac{8}{3}-2\log\frac{m_\tau^2}{m^{2}_{\tilde{\nu}_{\tau}}}
+\frac{m^2_{\tilde{\nu}_{\tau}}}{3m^2_{\tilde{\tau}_{R}}})\right],\nonumber\\
 G_{1}&= \frac{\alpha}{24\pi}\left[\frac{\lambda_{321}^{*}\lambda_{311}}
{m^{2}_{\tilde{\nu}_{\tau}}}(-\frac{8}{3}-2\log\frac{m^2_e}{m^{2}_{\tilde{\nu}_{\tau}}}
+\frac{m^2_{\tilde{\nu}_{\tau}}}{3m^2_{\tilde{e}_{R}}})\right.\\
&~~~~~~~~+\frac{\lambda_{322}^{*}\lambda_{312}}{m^{2}_{\tilde{\nu}_{\tau}}}
(-\frac{8}{3}-2\log\frac{m^2_\mu}{m^{2}_{\tilde{\nu}_{\tau}}}+
\frac{m^2_{\tilde{\nu}_{\tau}}}{3m^2_{\tilde{\mu}_{R}}})~~~~~\nonumber\\
&\left. ~~~~~~~~
+\frac{\lambda_{323}^{*}\lambda_{313}}{m^{2}_{\tilde{\nu}_{\tau}}}
(-\frac{8}{3}-2\log\frac{m_\tau^2}{m^{2}_{\tilde{\nu}_{\tau}}}
+\frac{m^2_{\tilde{\nu}_{\tau}}}{3m^2_{\tilde{\tau}_{R}}})\right],~~~~~\nonumber\\
A_{R}&=-\frac{\alpha
m_\mu}{48\pi}\frac{1}{m^{2}_{\tilde{\nu}_{\tau}}}
\left[\lambda_{321}^{*}\lambda_{311}
(1-\frac{m^2_{\tilde{\nu}_\tau}}{2m^2_{\tilde{e}_{R}}})
+\lambda_{322}^{*}\lambda_{312}(1-
\frac{m^2_{\tilde{\nu}_{\tau}}}{2m^2_{\tilde{\mu}_{R}}})+
\lambda_{323}^{*}\lambda_{313}(1-\frac{m^2_{\tilde{\nu}_{\tau}}}{2m^2_{\tilde{\tau}_{R}}})\right].
\end{align}
\end{subequations}
Notice that while the couplings $B_1$ and $B_2$ receive  a
contribution at a tree level, $A_R$ and $G_1$ receive
contributions only at the one-loop level.  If $\lambda_{ijk}$ are
the only sources of LFV, up to one-loop level, all other couplings
vanish. As discussed in the previous section,  the values of
$|B_1|^2$, $|B_2|^2$, $|A_L|^2$, $|A_R|^2$, $|G_1|^2+|C_2|^2/16$
and $|G_2|^2+|C_1|^2/16$ can be derived by combining information
from energy and angular distribution of the final particles in
$\mu \to eee$. Thus, the predicted pattern for the effective
couplings from $\lambda_{ijk}$ can be tested this way.

Let us now review the various bounds on the couplings from other
observations. At 1-loop level, the Yukawa couplings,
$\lambda_{ijk}$, contribute to the neutrino mass matrix, $m_\nu$
\cite{bound from neutrino mass matrix barbier} as well as to
processes such as $\tau \to \mu \nu \nu$ , $\tau \to e \nu \nu$
and $\mu \to e \nu \nu$ \cite{Barger:1989rktable,pdg}. The upper
bounds on the components of $(m_\nu)_{\alpha \beta}$ can be
translated into bounds on $\lambda$'s. The prediction of the SM
for charged lepton decay rates agrees with the measured values so
bounds can be set on the contribution from $\lambda$'s. In
particular, comparing the measured and calculated values of the
ratio $R_\tau\equiv {\rm \Gamma}(\tau \to e \bar{\nu}_e \nu_\tau)/
{\rm \Gamma}(\tau \to \mu \bar{\nu}_\mu \nu_\tau)$ gives bounds on
$\lambda$'s. In our analysis, we pick up values of $\lambda$ that
respect these bounds. The bounds that we use are summarized in
Table 1. The third column shows the observable from which the
bound is extracted. Notice that the bound on $\lambda_{12k}$ comes
from the numerical value of the CKM matrix. At first sight, this
might seem counterintuitive. Remember however that the value of
$V_{ud}$ is extracted by comparing $\Gamma(n \to p e \bar{\nu}_e)$
with $\Gamma(\mu \to e \bar{\nu}_e \nu_\mu)$. The unitarity
condition of the CKM matrix combined by the extracted values of
the CKM matrix elements yields a bound on the contribution from
$\lambda_{21k}$. For a full review of the bounds see
\cite{barbier}.
\begin{table}[t]
\begin{center}\begin{tabular}{|c|c|c|c|}\hline Coupling(s)&Bound&Observable&Ref\\\hline\hline
  $\lambda_{133}$&$9.4\times{10^{-4}}~\left(\frac{{m_{\tilde{\tau}_{R}}}}{100~{\rm GeV}}\right)^{1/2}$&
  $(m_\nu)_{ee}$&\cite{barbier}\\\hline
$\lambda_{233}$&$9.4\times{10^{-4}}~{\left(\frac{{m_{\tilde{\tau}_{R}}}}{100~{\rm
GeV}}\right)^{1/2}}$&$(m_\nu)_{\mu\mu}$& \cite{barbier}\\\hline
$\lambda_{i22}$&$1.5\times{10^{-2}}~{\left(\frac{{m_{\tilde{\mu}_{R}}}}{100~{\rm
GeV}}\right)^{1/2}}$&$(m_\nu)_{ii}
$&\cite{barbier}\\\hline$\lambda_{12k}$&$0.03~\frac{m_{\tilde{e}_{kR}}}{100~{\rm
GeV}}$&$V_{ud}$&\cite{Newbounds}
\\\hline $\lambda_{13k}$&$0.03~\frac{m_{\tilde{e}_{kR}}}{100~{\rm GeV}}$&$R_\tau$&\cite{Newbounds}\\\hline
$\lambda_{23k}$&$0.05~\frac{m_{\tilde{e}_{kR}}}{100~{\rm
GeV}}$&$R_\tau$&\cite{Newbounds}\\\hline$\lambda_{i23}\lambda_{j32}$
&$8\times{10^{-7}}~\left(\frac{m_{\tilde{\mu}_{R}}~m_{\tilde{\tau}_{R}}}{(100~{\rm
GeV)^2}}\right)^{1/2}$&$(m_\nu)_{ij}$&\cite{barbier}\\\hline
\end{tabular}\caption{Bounds on the trilinear $R$-parity violating couplings.
The masses of
${\tilde{e}_{R}}$, ${\tilde{\mu}_{R}}$ and ${\tilde{\tau}_{R}}$
are respectively indicated by
$m_{\tilde{e}_{1R}}$, $m_{\tilde{e}_{2R}}$ and
$m_{\tilde{e}_{3R}}$.
}\label{eatsource}\end{center}
\end{table}

\section{Transverse polarization and asymmetry}

Let us define P-odd asymmetry, $\mathcal{A}$, as \be
\mathcal{A}\equiv {\int_{0}^1 (d \Gamma/d\cos\theta)d\cos\theta-
\int_{-1}^0 (d \Gamma/d\cos\theta)d\cos\theta \over \int_{-1}^1 (d
\Gamma/d\cos\theta)d\cos\theta} \ . \label{MathcalA} \ee A nonzero
$\mathcal{A}$ violates parity. In fact, $\mathcal{A}$ is sensitive
to the P-odd combinations such as $|B_1|^2-|B_2|^2$.
 The  polarizations of
the final electron in $\mu^+\to e_1^+ e^-e_2^+$ can be defined as
\be \label{stplus} \langle s_{\hat{T}^-_i}\rangle\equiv {
\sum_{\vec{s}_{e^+_1},\vec{s}_{e^+_2}}\frac{ d\Gamma}{d \cos
\theta}|_{\vec{s}_{e^-}\cdot \hat{T}^-_i =|\vec{s}_{e^-}\cdot
\hat{T}^-_i|  }-\sum_{\vec{s}_{e^+_1},\vec{s}_{e^+_2}}\frac{
d\Gamma}{d \cos \theta}|_{\vec{s}_{e^-}\cdot \hat{T}^-_i
=-|\vec{s}_{e^-}\cdot \hat{T}^-_i|  }\over
\sum_{\vec{s}_{e^+_1},\vec{s}_{e^+_2},\vec{s}_{e^-}}\frac{
d\Gamma}{d \cos \theta} }\ \ee
 where $i=1,2,3$.
$\hat{T}_3^-$ is the longitudinal direction,
$\hat{T}_3^-=\vec{P}_{e^-}/|\vec{P}_{e^-}|$.
 $\hat{T}_1^-$ and $\hat{T}_2^-$ are unit vectors
 in the transverse directions defined as
\be \hat{T}_2\equiv {\vec{s}_\mu \times\vec P_{e^-}\over
|\vec{s}_\mu \times\vec P_{e^-}|} \ \ {\rm and}  \ \
\hat{T}_1\equiv {\hat{T}_2\times\vec P_{e^-}\over
|\hat{T}_2\times\vec P_{e^-}|}. \label{directions}\ee Finally,
 $\theta$ is the angle between the momentum of $e^-$
   and the polarization of the muon:
   $\theta=\arccos [\vec{s}_\mu\cdot\vec P_{e^-}/
   (|\vec{s}_\mu|\cdot|\vec P_{e^-}|)]$ .

As shown in \cite{eee}, while $\langle s_{\hat{T}_3^-}\rangle$ is
sensitive only to the absolute values of the couplings, the
transverse polarizations $\langle s_{\hat{T}_1^-}\rangle$ and
$\langle s_{\hat{T}_2^-}\rangle$ are sensitive to the CP-violating
phases in the effective Lagrangian.
 Straightforward but cumbersome
calculation shows that within the present model with the coupling
pattern in Eqs. (\ref{Cs}),
$$ \langle s_{T_1^-}\rangle \propto \Re[ B_1B_2^*-12 A_RB_1^*]$$
and
$$ \langle s_{T_2^-}\rangle \propto \Im[ B_1B_2^*-12 A_RB_1^*]\ .$$
 Of course, similar
formulas hold for the transverse polarization of the final
positron in $\mu^- \to e^- e^+ e^-$.

\begin{figure}[t!]
\begin{center}
\includegraphics[height=6 in,width=8 in,angle=0]{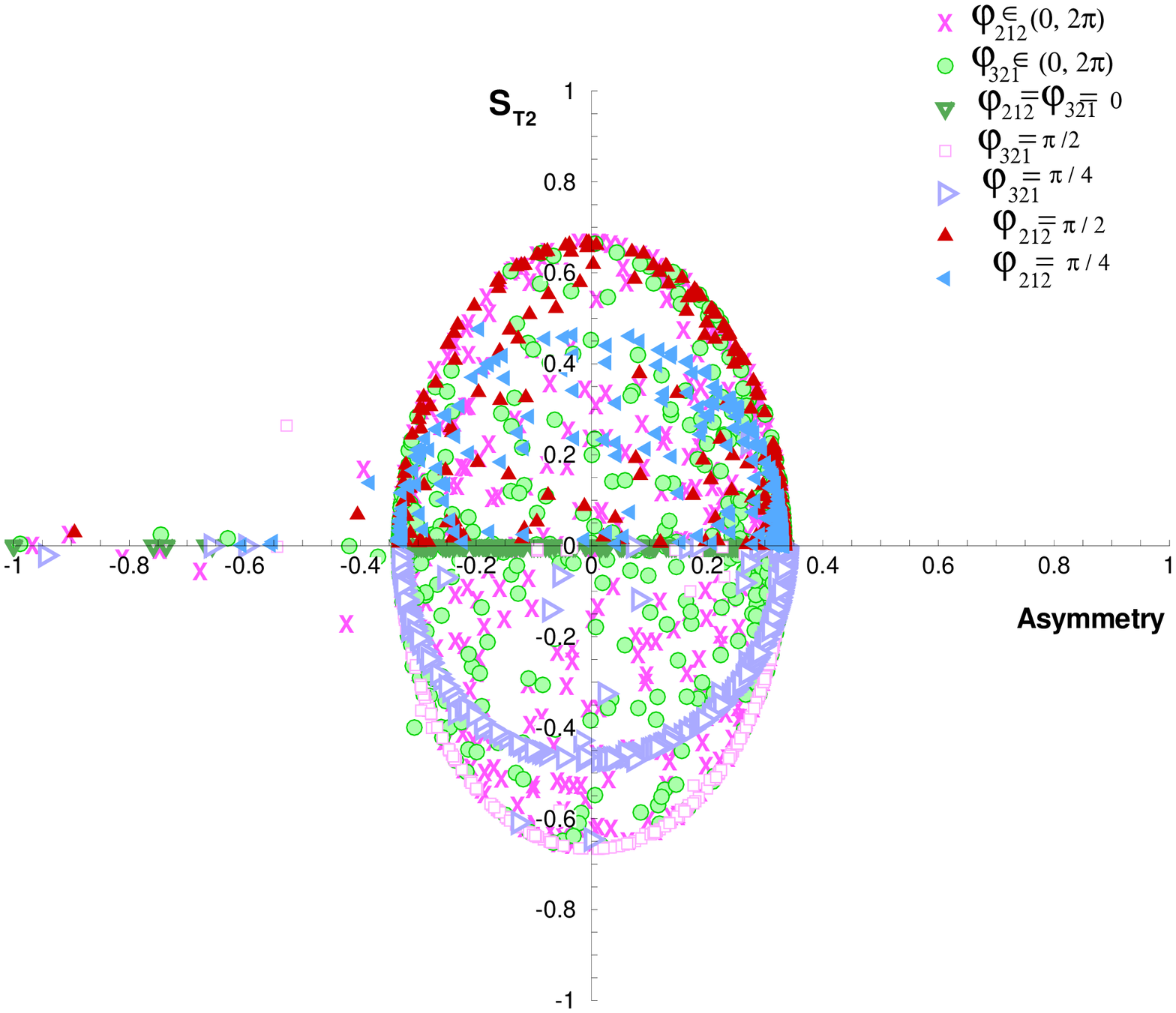}
\caption{Transverse polarization of the  electron in $\mu^+ \to
e^+ e^-e^+$  versus the P-odd asymmetry $\mathcal{A}$ defined in
Eq.~(\ref{MathcalA}).  The input values for masses correspond to
the ones at the $\alpha$ benchmark in \cite{benchmark} with $
m_{\tilde{\nu}_\mu} \simeq m_{\tilde{\nu}_\tau}\simeq 285$ GeV.
Random values between $10^{-5}$ up to bounds in Table 1 are
assigned to the $\lambda$ couplings and points at which ${\rm
Br}(\mu \to eee)\in 5 \cdot 10^{-13}(1\pm 10\%)$ are selected.
 To calculate $\langle
s_{\hat{T}_2}\rangle$ and $\mathcal{A}$, we have set
$\mathbb{P}_\mu =100\%$ and $\theta=\pi/2$ (see definitions in
Eqs.~(\ref{stplus},\ref{directions})). $\varphi_{ijk}$  is the
phase of $\lambda_{ijk}$. Points with different colors and symbols
correspond to a nonzero value for $\varphi_{ijk}$ as described in
the legend. For each set, the rest of phases vanish. For points
shown by $\times$ and $\circ$, the corresponding phase takes
random values at a linear scale from ($0,2\pi$). }
\label{pointsoveregg}\end{center}
\end{figure}
\
\begin{figure}[t!]
\begin{center}
\includegraphics[height=6 in,width=8 in,angle=0]{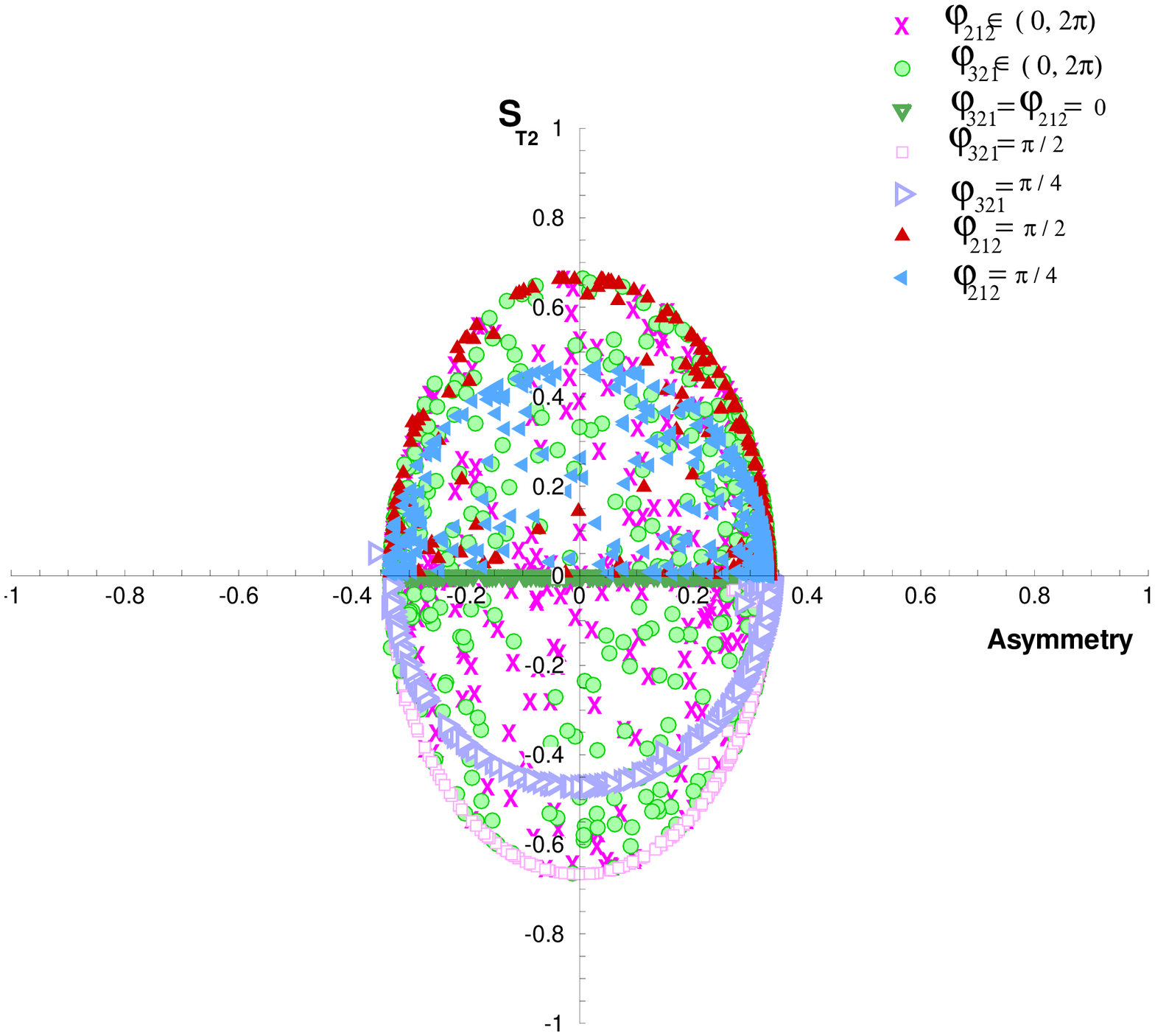}
\caption{The same as Fig.~\ref{pointsoveregg}, except that we have
removed the points at which $8|G_1|^2/(|B_1|^2+|B_2|^2)>0.05$. }
\label{perfect egg}
\end{center}
\end{figure}
Let us suppose ${\rm Br}(\mu \to e ee)$ is measured and found to
be close to $10^{-12}$. Let us moreover suppose that the Dalitz
plots
 reveal that $B_1$ and $B_2$ give the dominant contribution to
this decay as it is expected in the case that LFV effects
originate from $\lambda_{ijk}$. If the MEG experiment at PSI
reports a null result ({\it i.e.,} ${\rm Br}(\mu \to
e\gamma)<10^{-13}\ $), within the context of MSSM (to be tested at
the LHC), such a set of conditions attests our assumption that
$\lambda_{ijk}$ is the prime source for LFV. If MEG also finds a
signal for $\mu \to e \gamma$, other LFV  terms, such as slepton
masses or the trilinear soft supersymmetry breaking terms, might
contribute to Br$(\mu \to e\gamma)$ but as we discussed earlier,
even in this case, we can neglect the contribution from $A_L$ and
$A_R$ to Br($\mu \to eee)$.  LFV terms in slepton masses or the
trilinear soft supersymmetry breaking terms can contribute to
other effective couplings in (\ref{effectiveLag}) but the effect
will be loop suppressed and negligible.
 Notice that
although within the model under our study, $G_1$ and $A_R$ are
given by the same combinations of $\lambda_{ijk}$, $G_1$ is
enhanced by a factor of $\log (m_{\tilde{\nu}}^2/m_l^2)$ so we do
not neglect its contribution  [see Eq.~(\ref{Cs})].
 Neglecting the effects of $A_R$, the formulas for
$\mathcal{A}$, $\langle s_{\hat{T}^-_1} \rangle $ and $\langle
s_{\hat{T}^-_2} \rangle $ will have forms: \be \label{Aformula}
\mathcal{A}={|B_2|^2-|B_1|^2-24|G_1|^2 \over
3(|B_1|^2+|B_2|^2+8|G_1|^2)} \mathbb{P}_\mu \  \ee  and
\begin{subequations} \label{Ss}
  \begin{align}
  \langle s_{\hat{T}_1^-}\rangle &=
   {4 {\Re}[B_1B_2^*] \mathbb{P}_\mu \sin\theta
  \over |B_1|^2(3- \mathbb{P}_\mu
  \cos \theta)+|B_2|^2(3+ \mathbb{P}_\mu\cos \theta)+24|G_1|^2(1- \mathbb{P}_\mu\cos \theta)},\\
 \langle s_{\hat{T}_2^-}\rangle &= {4 {\Im}[B_1B_2^*]  \mathbb{P}_\mu\sin\theta
  \over |B_1|^2(3- \mathbb{P}_\mu\cos \theta)+|B_2|^2(3+ \mathbb{P}_\mu\cos \theta)+24|G_1|^2(1- \mathbb{P}_\mu\cos \theta)}\
  ,
\end{align}
\end{subequations}
in which $ \mathbb{P}_\mu$ is the polarization of the initial
muons.

 Moreover the longitudinal polarization is given by
$$\langle s_{\hat{T}_3^-}\rangle=
{|B_1|^2-|B_2|^2+8|G_1|^2- \mathbb{P}_\mu\cos
\theta\left[ (|B_1|^2+|B_2|^2)/3+8|G_1|^2 \right] \over
|B_1|^2+|B_2|^2+8|G_1|^2+ \mathbb{P}_\mu\cos \theta\left[
(|B_2|^2-|B_1|^2)/3-8|G_1|^2 \right]} \ . $$

 Notice when the
electron is emitted in the direction perpendicular to the spin of
the muon, the transverse polarization is maximal. In our analysis,
we will set $\theta=\pi/2$ which experimentally means we study the
data from the polarimeter collecting electrons with momentum
perpendicular to $\vec{s}_\mu$. If more than a single polarimeter
is installed, more data can of course be collected. The average
polarization can be defined as \be \overline{\langle
s_{\hat{T}_i^-} \rangle} \equiv {\int_{-1}^1\langle
s_{\hat{T}_i^-} \rangle~ [d \Gamma(\mu\to eee)/d\cos \theta]~d
\cos \theta \over \int_{-1}^1 [d \Gamma(\mu\to eee)/d\cos
\theta]~d \cos \theta} . \ee For $i=1,2$, $ \overline{\langle
s_{\hat{T}_i^-} \rangle} = \frac{\pi}{4} \mathbb{P}_\mu \langle
s_{\hat{T}_i^-} \rangle|_{ \theta =\frac{\pi}{2}}.$ It is
noteworthy that if the values of Br$(\mu \to eee)$, $\mathcal{A}$
and $\langle s_{\hat{T}^-_3} \rangle|_{ \theta =\frac{\pi}{2}}$
are measured, the numerical values of $|B_1|$, $|B_2|$ and $|G_1|$
can be derived.


\section{Feasibility of measuring the transverse polarization}

The typical experimental setups devoted to the study of muon decay
(such as the MEG experiment or the experiment at TRIUMF described
in \cite{triumf}) can be summarized as follows. A proton beam
collides on a target producing pions. Charged pions are stopped in
the target and decay at rest into a neutrino and a muon. Muons at
production are 100 \% polarized up to negligible correction due to
the neutrino mass \cite{triumf}. Muons exit the first target and
are transmitted to a second target where they stop and then decay
at rest. Based on the setup of the experiment, muons of either
positive or negative sign can be selected to be transmitted to the
second target. Negative muons would form bound states with atoms
in the second target so we focus on the decay of positive muons
which decay as free states. When muons decay, they are still
highly polarized. The degree of depolarization from the production
to decay depends on the setup of experiment. Especially if only
the muons produced at the surface of the first target are
collected (as it is done both in MEG \cite{Megsite} and in the
experiment described in \cite{triumf}), the depolarization will be
quite negligible. For example, in the TRIUMF experiment described
in \cite{triumf}, the polarization remains above 99\% until the
muon decay in the second target. For the purpose of the present
analysis we can safely replace $\mathbb{P}_\mu=100\%$. Notice that
unlike the case of \cite{triumf}, even a moderate accuracy in
knowledge of $\mathbb{P}_\mu$ will be sufficient to perform the
present analysis so from this aspect, it is easier to carry out
the present measurement \cite{triumf}.

Let us now discuss the possibility of measuring the transverse
polarization of the electron in $\mu^+ \to e^+ e^-e^+$. First, let
us recall  the measurement of the transverse polarization of the
positron in $\mu^+\to e^+\nu_e\bar{\nu}_\mu$ which was carried out
about 25 years ago to extract the Michel parameters
\cite{Burkard:1985wn}.  The outline of the polarization
measurement is as follows. The emitted positrons interact with
electrons in a target which  are polarized in a direction
perpendicular to the momentum of the emitted positrons by a
magnetic field. The electron positron pair annihilate into a
photon pair. By studying the azimuthal distribution of the final
photon pairs, the transverse polarization of the emitted positron
can be determined. In our case, we are interested in measuring the
transverse polarization of the final electron rather than the
final positron. Fortunately, a similar setup can be employed to
determine the transverse polarization of the emitted electron. Of
course in this case instead of annihilation into a photon pair,
the M\"oller scattering ($e_1^- e_2^- \to e_3^- e_4^-$) will take
place. Similar to the case of $e^-e^+\to \gamma \gamma$, the
azimuthal distribution of the final particles carry information on
the spin of the emitted  electron. Let us take the $\hat{z}$
direction to be parallel to the momentum of the emitted electron
(the momentum of $e_1^-$) and take $\hat{x}$ to be in the
direction of the spin of the electron at rest in the magnetized
target ({\it i.e.}, the spin of $e_2^-$):
$$P_1=(\sqrt{k^2+m_e^2},0,0,k) \ \ \ {\rm and} \ \ \
P_2=(m_e,0,0,0)\ . $$ The spin of $e_1$ can be described by $(a \
b)$ where $|a|^2+|b|^2=1$: $$\langle s_x\rangle=2 \Re[ba^*], \
\langle s_y\rangle=2 \Im[ba^*] \ {\rm and} \ \langle
s_z\rangle=|a|^2-|b|^2\ . $$ The angular distribution of the final
particles are described by $(\theta, \phi)$:
$$P_3=(\sqrt{k'^2+m_e^2},k'\sin\theta \cos\phi, k'\sin \theta \sin
\phi,k'\cos \theta) \ {\rm and} \ P_4=P_1+P_2-P_3 \ ,$$ where the
energy-momentum conservation implies
$$k'={m_e k \over m_e +k(1-\cos \theta)} \ . $$  A cumbersome but
straightforward calculation shows that \be \int_0^{2\pi} \frac{d
\sigma}{d \cos \theta~d \phi} \cos \phi d \phi=\ee
$$(|b|^2-|a|^2){k'^4e^4(1-\cos \theta)\sin \theta\over 64\pi
m_e^4(k-k')}\left( {k'(1-\cos \theta)\over (k-k')^2}+{k'(1-\cos
\theta) -m_e \over (k-m_e)^2}-{m_e (1-2k'\cos \theta/k) \over
(k-k')(k-m_e)} \right)$$ and \be \int_0^{2\pi} \frac{d \sigma}{d
\cos \theta~ d \phi} \sin 2 \phi d \phi=\ee
$$\Im[ab^*]{k'^5e^4(1-\cos \theta)\sin^2 \theta\over 64\pi
m_e^4(k-k')}\left( {m_e/k\over (k-k')^2}+{m_e/k \over
(k-m_e)^2}-{1 \over (k-k')(k-m_e)} \right)\ .
$$
Thus, by measuring the partial cross section, $d
\sigma/(d\cos\theta~d\phi)$, and taking the above integrals, $a$
and $b$ (up to an overall phase) can be determined and the spin of
$e_1$ can be therefore reconstructed.

There are two problems that complicate the measurement: (1) In the
lab frame where $e^-_2$ is at rest, the majority of the electrons
are scattered in the forward direction within a narrow cone with
opening angle of $O(2m_e/k)$ where $k\sim m_\mu /3\sim 30~{\rm
MeV}$. The same problem existed in the case of measuring the
polarization of the emitted positron. (2) The scattered electron
$e_3$ can again scatter on the electrons in the magnetized target
before exiting it. Multiple scattering will distort the  azimuthal
distribution in which the information of the spin of the initial
electron is imprinted. The same problem existed is the case of
measuring the spin of the positron as the photons produced in the
electron positron annihilation could Compton scatter on the
electrons in the magnetized target. In both cases, the total
scattering cross section is of order of $e^4/(16\pi m_e k)$.
Fortunately, there are established techniques to overcome these
difficulties.

\section{Combined analysis of $\mathcal{A}$ and $\langle
s_{\hat{T}_2^-}\rangle$}   By rephasing the leptonic fields, three
 out of nine phases in the  $\lambda$ couplings  can be absorbed.
 Let us consider the
basis in which $\lambda_{311}$, $\lambda_{211}$ and
$\lambda_{312}$ are all real. The rest of $\lambda$'s in this
basis can in general be complex among which the phases of
$\lambda_{321}$, $\lambda_{212}$, $\lambda_{322}$ and
$\lambda_{323}$ can lead to a nonzero $\langle
s_{\hat{T}^-_2}\rangle$. We  first concentrate on the phases of
$\lambda_{321}$ and $\lambda_{212}$ which contribute to $B_1$ at a
tree level. We then comment on the rest of phases.
 As we discussed in the previous section,  we can safely replace
 $\mathbb{P}_\mu=100~\%$ and that is what we do in the following.  For any given polarization, our
results can be simply rescaled.

Figs.~1 and 2 demonstrate the dependence of $\langle
s_{\hat{T}^-_2}\rangle$ and $\mathcal{A}$ on the phases. To draw
these scatter plots, we have assigned random numbers at a
logarithmic scale from $10^{-5}$ to the upper bound on
$|\lambda|$'s. ($|\lambda|$'s take up random values in a
logarithmic scale.) As explained in the caption, various values
are assigned to the phase of either $\lambda_{321}$ or
$\lambda_{212}$, setting
 the other one (as well as the rest of phases) equal to zero. The
 phase of $\lambda_{ijk}$ is denoted by $\varphi_{ijk}$.
We have selected the configurations of $\lambda$ for which Br$(\mu
\to eee)$ lies within the range Br$(\mu \to eee)=5\times
10^{-13}(1 \pm 10\%) $. From an experimental perspective, this
means that we have supposed Br$(\mu \to eee)$ is measured to be in
the range $5\times 10^{-13}(1 \pm 10\%) $. The 10\% uncertainty is
a nominal value that we have taken as an example to highlight the
fact that the branching ratio measurement will
 suffer from a finite uncertainty. Our results are robust against varying
 the value of this uncertainty. In fact by rescaling the coupling by
 a $\delta N$ percent, Br$(\mu \to eee)$ will change by $4 \delta N \%$ but
  $\langle s_{\hat{T}_2^-}\rangle$ and  $\mathcal{A}$, being
 ratios, will not vary.

   Points denoted by pink crosses
are the points at which $\varphi_{321}=0 $  and $\varphi_{212}$
takes up random values in $(0,2\pi)$. On the contrary, those shown
by green circles correspond to the cases that $\varphi_{212}=0$
and $\varphi_{321}$ takes random values in the range $(0,2\pi)$.
As seen from the figures, the areas over which pink $\times$ and
green $\circ$ are scattered completely overlap which means if
$|\lambda|$'s are unknown, the two solutions are
indistinguishable. However, valuable information from $\langle
s_{\hat{T}_2^-}\rangle $ can be derived. For example if both
CP-violating phases vanish, $\langle s_{\hat{T}_2^-}\rangle $ also
vanishes (points indicated by green $\nabla$).

Remember that, while  $B_1$ and $B_2$ receive nonzero
contributions at a tree level, $G_1$ receives a contribution only
at a loop level. As long as $G_1$ has a negligible value, for
given $\mathcal{A}$ and Br($\mu \to eee)$, $|B_1|$ and $|B_2|$ are
fixed. It is straightforward to check that, for $|G_1|\to 0$, \be
|\mathcal{A}|<\frac{1}{3} \ \ \ {\rm and} \ \ \ |\langle
s_{\hat{T}_2^-}\rangle |<\frac{2}{3}\sqrt{1-9 \mathcal{A}^2} \ .
\label{boundaries} \ee As seen in the figures, the majority of
points lie inside an oval-shaped region whose boundaries are given
by Eq.~(\ref{boundaries}). These are the points for which the
contribution from loop suppressed $|G_1|^2$ can be neglected.
 As seen from  Fig.~1, there are only few points (about
2 percent of all points) lying outside the oval-shaped region. At
points with $\mathcal{A}<-1/3$, the contribution from $|G_1|^2$
dominates.
 For $\mathcal{A}<-0.5$, we find $|\langle
 s_{\hat{T}_1^-}\rangle|,|\langle
 s_{\hat{T}_2^-}\rangle|\ll 0.1$ which is expected from Eqs.
 (\ref{Aformula}) and (\ref{Ss}).
In Fig.~2, we have removed the points for which
$8|G_1|^2/(|B_1|^2+|B_2|^2)>0.05$. As a result, Fig.~2 does not
include points outside the oval-shaped region. If the pair
$(\mathcal{A},\langle s_{\hat{T}_2^-}\rangle$) turns out to be
outside the oval-shaped region, it means $|G_1|$ is relatively
large.  This can happen if $\lambda_{322}$ is more than  50 times
larger than the rest of $\lambda_{ijk}$ (see Eqs. (\ref{Cs}) and
table 1) so $\mathcal{A}<-1/3$ indicates that the flavor structure
of the $\lambda_{ijk}$ coupling should be hierarchical.

 As mentioned above, in the  limit $G_1\to 0$, for a given
 $\mathcal{A}$ and $Br(\mu \to eee)$, $|B_1|$ and $|B_2|$ are
 fixed so $\langle s_{\hat{T}_2^-} \rangle$ determines $\arg[B_2
 B_1^*]$. Thus, if all the phases except $\varphi_{321}$ are zero,
$\langle s_{\hat{T}_2^-} \rangle$ will determine $\varphi_{321}$.
 This can be seen in Fig.~2. That is impressive that $\varphi_{321}$ can be
 derived  even without knowledge of $|\lambda_{ijk}|$'s  (of course under the
 assumption of a single nonzero phase).
Deriving the value of $\varphi_{212}$ is going to be more
challenging even when we set all the other phases equal to zero.
As seen from Figs.~1 and 2, for a given  $\mathcal{A}$ and a
certain value of $\varphi_{212}$, depending on the configuration
of the absolute values of the $\lambda$'s, $\langle
s_{\hat{T}_2^-} \rangle$ can take any value between zero and its
maximal value which is $(2/3) (1-9\mathcal{A})^{1/2}\sin
\varphi_{212}$.
%
This is understandable as $B_2$ receives contributions from
various terms (compare Eqs. (\ref{Cb}) and (\ref{Cc})) so unlike
the case of ($\varphi_{212}=0$, $\varphi_{321}\ne 0$), in this
case, the nonzero phase is not given by $\arg[B_1 B_2^*]$.

From Fig.~2, we observe that for $|\mathcal{A}|<0.2$, by
simultaneous measurements of $\mathcal{A}$ and $\langle
s_{\hat{T}_2^-} \rangle$ with reasonable accuracy, even without
independent knowledge of $|\lambda_{ijk}|$, solutions with
($\varphi_{321}=\pi/2$,$\varphi_{212}=0$) and
($\varphi_{321}=\pi/4$,$\varphi_{212}=0$) can be distinguished
(see points denoted by violet $\triangleright$ and pink
$\square$). Notice that all points denoted by red $\vartriangle$
and blue $\triangleleft$  corresponding respectively to
($\varphi_{321}=0$,$\varphi_{212}=\pi/2$) and
($\varphi_{321}=0$,$\varphi_{212}=\pi/4$) lie above the horizontal
axis. Solutions with positive  and negative $\sin \varphi_{212}$
are distinguishable but deriving the value of $\varphi_{212}$
without knowledge of $|\lambda_{ijk}|$ does not seem to be
practical. We have found that 
 the contributions  from the
phases that enter only at loop level ({\it i.e.,} $\varphi_{322}$,
$\varphi_{313}$ and $\varphi_{323}$) to $\langle
s_{\hat{T}^-_2}\rangle$ are smaller than $0.1$. Thus, in deriving
the values of $\varphi_{212}$ and $\varphi_{321}$ from $\langle
s_{\hat{T}_2^-} \rangle$, the potential contributions from the
rest of the phases can be ignored.

As  discussed above, an upper bound on $|G_1|$ can considerably
simplify the analysis and solve the degeneracies.  Although
 $|G_1|^2$ (more precisely, $|G_1|^2+|C_2|^2/16$) can  in principle be extracted from  the energy
 distribution of final particles, within the present model, its
 value will most probably be too small to be measured so in practice   only an
 upper bound on $|G_1|$ can be extracted as we have assumed in deriving
 Fig.~2.  Notice that
 $$\lim_{ ~~~~|G_1|\to 0}  \  \frac{\langle s_{\hat{T}_3^-} \rangle
|_{\theta=\frac{\pi}{2}} }{\mathcal{A}}=3\ .$$ That is while if
$G_1$ and $G_2$ (or $C_1$ and $C_2$) gave the main contribution to
$\mu \to eee$, we would expect that $\langle s_{\hat{T}_3^-}
\rangle |_{\theta=\frac{\pi}{2}} /\mathcal{A}=1$. The ratio of
longitudinal polarization to $\mathcal{A}$ can therefore be
regarded as a cross-check for the smallness of $|G_1|$.

 To draw Figs~1 and 2, we have used the spectrum at the
$\alpha$ benchmark \cite{benchmark} as the input: {\it i.e.,} We
have set $m_{\tilde{\nu}_\mu}=m_{\tilde{\nu}_\tau}=285~{\rm GeV}
$. For two reasons, we expect the results to be robust against
varying the input masses: (i) Varying $m_{\tilde{\nu}_\mu}^2$ and
$m_{\tilde{\nu}_\tau}^2$ is respectively equivalent to rescaling
$\lambda_{211}$ and $\lambda_{311}$ (see Eqs. (\ref{Cb}) and
(\ref{Cc})). (ii) Both $\mathcal{A}$ and $\langle s_{\hat{T}_2^-}
\rangle$ are defined as ratios so the dependence on the
supersymmetry scale disappears. We have re-drawn the diagram for
different benchmarks. As expected, the results are not sensitive
to the input for the mass spectrum.

It is noteworthy that if the only sources of LFV are the
$\lambda_{ijk}$'s giving rise to Br$(\mu \to eee)$, Br$(\mu \to
e\gamma)$ will be smaller than $10^{-13}$ so if the MEG experiment
reports a $\mu \to e \gamma$ signal, sources for $\mu \to e
\gamma$ other than $\lambda_{ijk}$'s must exist.

As seen in Figs.~1 and 2, both $\mathcal{A}$ and $\langle
s_{\hat{T}_2^-} \rangle$ can be relatively large so as long as the
errors in their measurement ({\it i.e.,} $\delta \mathcal{A}$ and
$\delta\langle s_{\hat{T}_2^-} \rangle$) are below $\sim 0.1$,
their nonzero values can be established. Suppose the numbers of
electrons studied to derive $\mathcal{A}$ and $\langle
s_{\hat{T}_2^-} \rangle$ are respectively $\mathcal{N}_A$ and
$\mathcal{N}_s$. We roughly expect the statistical errors to be
$\delta \mathcal{A}\sim 1/\sqrt{\mathcal{N}_A}$ and $\delta
\langle s_{\hat{T}_2^-} \rangle\sim 1/\sqrt{\mathcal{N}_s}$. To
derive $\mathcal{A}$, the majority of the emitted electrons can in
principle be employed, so with a few hundred $\mu^+\to e^+e^-e^+$
decays, the statistical error in the measurement of $\mathcal{A}$
will be reasonably small and below 0.1.  However, we expect only a
fraction of the emitted electrons, $r$,  to enter the polarimeters
so for establishing nonzero $\langle s_{\hat{T}_2^-} \rangle$, the
total number of $\mu^+\to e^+e^-e^+$ decays has to be larger than
$100/r$. That is if $r\sim 10 \%$, the total number of $\mu^+ \to
e^+e^-e^+$ has to be larger than a few thousand.

In the above analysis, we have employed information on the
$R$-parity violating couplings from only the LFV rare decays. The
R-parity violating couplings can  in principle be directly
measured by accelerators. $|\lambda_{i11}|$ leads to a resonant
production of $\tilde{\nu}_i$ in a $e^-e^+$ collider ($e^-e^+\to
\tilde{\nu}_i$) so  $|\lambda_{i11}|$ can be derived provided that
$|\lambda_{i11}|>10^{-5}$ and the center of mass energy is equal
to the mass of $\tilde{\nu}_i$ \cite{sneutrinodecay,barbier}.
Moreover, the $\lambda$ couplings can lead to $\tilde{\nu}_i \to
l_j^+l_i^-$, $\tilde{\chi}_1^0 \to \bar{\nu}_i l_j^+l_k^-$ and
$\tilde{\chi}_1^+\to \nu_k\bar{\nu}_i l_j^+,l_i^+l_k^+l_j^-$
\cite{Barbier:1998fe,barbier} where $\tilde{\chi}_1^0 $ and
$\tilde{\chi}_1^+$ are the lightest neutralino and chargino. Thus,
by measuring the decay length and the flavor of the final charged
leptons, $|\lambda_{ijk}|$ can in principle be extracted. If
$|\lambda_{ijk}|<\mathcal{O}(10^{-5})$, the decay length will be
too small to be resolved \cite{barbier}. This method is therefore
sensitive only to the values of $|\lambda_{ijk}|$ smaller than
$\mathcal{O}(10^{-5})$.\footnote{Even if the decay length is not
resolved, a combination of $|\lambda|$'s may be extracted. For
example, consider chain processes $e^-e^+\to Z^*\to \tilde{\nu}_i
\bar{\tilde{\nu}}_i$ and the subsequent decays $\tilde{\nu}_i\to
e^- \mu^+$ and $\bar{\tilde{\nu}}_i\to \mu^- \mu^+$. Such a chain,
being LFV, is not contaminated by the SM or $R$-parity conserving
MSSM so even if the decay lengths of $\tilde{\nu}_i\to e^- \mu^+$
and $\bar{\tilde{\nu}}_i\to \mu^- \mu^+$ are too short to be
resolved, the possibility of deriving information on the relevant
couplings is still open. Considering all such possibilities is
beyond the scope of the present letter.} If  $|\lambda|$'s are all
smaller than $10^{-5}$, each $|\lambda_{ijk}|$ might be extracted
from $\tilde{\nu}_i\to l_j^+l_i^-$, $\tilde{\chi}_1^0 \to
\bar{\nu}_i l_j^+l_k^-$ and $\tilde{\chi}_1^+\to \nu_k\bar{\nu}_i
l_j^+,l_i^+l_k^+l_j^-$ but in this case $Br(\mu \to eee)$ will be
too small ($Br(\mu \to eee)<10^{-16}$). Let us now consider the
range, $m_{susy}\sim 100$~GeV,
$\lambda_{211},\lambda_{311}>10^{-3}$ and $\lambda_{ijk}\sim
10^{-5}$ with $ijk\ne 211,311$. In this range, $e^-e^+\to
\tilde{\nu}_i$ yields $|\lambda_{i11}|$ and Br$(\mu \to eee)$ is
close to the present bound. Moreover, for $ijk\ne 211,311$,
$\tilde{\nu}_i\to l_j^+l_k^-$ and $\tilde{\chi}_1^0 \to
\bar{\nu}_i l_j^+l_k^-$ will have a resolvable decay length but
the point is that the decay modes involving $\lambda_{211}$ and
$\lambda_{311}$ will dominate and lead to a decay length too short
to be observable: e.g., $\Gamma(\tilde{\nu}_\mu\to
e^+e^-)/\Gamma(\tilde{\nu}_\mu\to \tau^+\tau^-)\sim {10^4}$. Thus,
even in case that the $R$-parity conserving decay modes are
kinematically forbidden, extracting the decay lengths  will be
quite challenging. Let us however suppose that these experimental
difficulties are partly solved and certain $|\lambda_{ijk}|$ (but
not necessarily all) are measured. Such achievement might not be
out of reach  if $\lambda_{i11} \sim 10^{-4}$ and the rest of
$\lambda$'s are of order of $10^{-5}$. Complementary information
can then be derived from Br($\mu \to eee$), $\mathcal{A}$ and
$\langle s_{T_2^-}\rangle $: Neglecting the $|G_1|^2$ effects,
Br($\mu \to eee$) and $\mathcal{A}$ give $|B_1|$ and $|B_2|$ which
to leading order correspond to
$|\lambda_{311}|\cdot|\lambda_{321}|$ and $|
\lambda_{211}^*\lambda_{212}+(m_{\tilde{\nu}_\mu}^2/
m_{\tilde{\nu}_\mu}^2)\lambda_{311}^*\lambda_{312}|$,
respectively. Thus, if $\lambda_{311}$ is extracted from $e^+ e^-
\to \tilde{\nu}_\tau$ at ILC, the measurement of  Br($\mu \to
eee$) and $\mathcal{A}$ gives $|\lambda_{321}|$. If
$|\lambda_{211}|$, $|\lambda_{212}|$, $|\lambda_{311}|$ and
$|\lambda_{312}|$ are all derived by accelerators, this method
will give the phase of $\lambda_{212}$. The measurement of
$\langle s_{\hat{T}_2^-}\rangle$ will then yield the phase of
$\lambda_{321}$.

\section{Conclusions and discussion}
The trilinear $R$-parity violating Yukawa couplings,
$\lambda_{ijk}\hat{L}_i\hat{L}_j \hat{E}_k/2$ can lead to $\mu \to
e ee$. In particular, $\lambda_{321}$, $\lambda_{311}$,
$\lambda_{211}$, $\lambda_{312}$ and $\lambda_{212}$ contribute to
$\mu \to e ee$ at tree level. By rephasing the lepton fields, we
can go to a basis in which $\lambda_{311}$, $\lambda_{312}$ and
$\lambda_{211}$ are real. This exhausts the freedom to rephase the
other fields so $\lambda_{321}$ and $\lambda_{212}$ can in general
be complex and can be considered as sources for CP-violation.
Their phases can induce transverse polarization for $e^-$ in the
direction $\hat{T}_2^-=\vec{s}_\mu \times
\vec{P}_{e^-}/|\vec{s}_\mu \times \vec{P}_{e^-}|$. Thus, by
measuring this polarization, one can derive information on the
CP-violating phases. We have shown that for maximal CP-violation,
$|\langle s_{\hat{T}_2^-} \rangle|$ can reach 2/3 so with a
moderate sensitivity to $\langle s_{\hat{T}_2^-} \rangle$,
CP-violation can be established. The sign of the CP-violating
phase can also be determined by measuring the sign of $\langle
s_{\hat{T}_2^-}\rangle$.

We have also studied the $P$-odd asymmetry, $\mathcal{A}$ defined
in Eq.~(\ref{MathcalA}) and discussed the information that from a
combined analysis of $\mathcal{A}$ and $\langle s_{\hat{T}_2^-}
\rangle$ can be obtained. For the majority of the $\lambda$
configurations, the tree level effects dominate so the effective
coupling $|G_1|$ is much smaller than $|B_1|$ and $|B_2|$ and its
effects can therefore be neglected. In this case, $|G_1|$ will be
too small to be measured but an upper bound can be put on $|G_1|$
by studying the energy distribution of the final particles in $\mu
\to eee$ or as discussed in the present paper by combining
information on $\mathcal{A}$, Br($\mu \to eee$) and $\langle
s_{\hat{T}_3^-} \rangle$. We have noticed that if
$8|G_1|^2<0.05(|B_1|^2+|B_2|^2)$, the analysis becomes much
simpler because of two reasons: (i) The contributions of the
phases of $\lambda_{322}$, $\lambda_{323}$ and $\lambda_{313}$ to
$\langle s_{\hat{T}_2^-} \rangle$ can be neglected. These are the
couplings that contribute only at a loop level. (ii) For given
$\mathcal{A}$ and Br($\mu \to eee$), the absolute values of $B_1$
and $B_2$ are fixed which simplifies  the analysis. In particular,
restricting the analysis to a single CP-violating phase, the
simultaneous measurement of $\mathcal{A}$ and $\langle
s_{\hat{T}_2^-}\rangle$ yields the phase of $\lambda_{321}$ even
if the values of $|\lambda_{ijk}|$ are not a priori known.
Degeneracies however exist between solutions for which  the phases
of $\lambda_{212}$ and $\lambda_{321}$ are both nonzero. By
measuring $\mathcal{A}$ and $\langle s_{\hat{T}_2^-}\rangle$,
different classes of solutions can be distinguished. If the
absolute values of $\lambda$ are measured by an accelerator-based
experiment (or by some other methods), the measurements in $\mu
\to eee$ can yield both phases. In fact, if the tree level
contribution dominates, the relevant $\lambda$ parameters can be
over-constrained.

 Neglecting the contribution of $|G_1|$,
we find $-1/3<\mathcal{A}<1/3$ and $-2\sqrt{(1-9
\mathcal{A}^2)/9}<\langle s_{\hat{T}_2^-}\rangle< 2\sqrt{(1-9
\mathcal{A}^2)/9}$.
 Only at a small fraction of the parameter
space where ($\lambda_{322}\gg$ rest of $\lambda$), the loop
effects can dominate and  lead to $\mathcal{A}<-1/3$. Within the
model under consideration, $\mathcal{A}<-1/3$ therefore indicates
a hierarchical flavor structure for the $\lambda$ parameters.

Stopped $\mu^-$ would form bound states with the atoms before they
decay. Because of this technical difficulty, we have concentrated
on the decay of $\mu^+$ and the spin of the final electron emitted
from it.
  Similar consideration holds for the positron
 in free decay of negative muon; {\it i.e.,} for $e^+$ in
 $\mu^- \to e^- e^+e^-$. Within the present model, the
  transverse polarizations of the  electrons in $\mu^- \to e^- e^+e^-$
  (or that of the positrons in $\mu^+ \to e^+ e^-e^+$)
  are loop-suppressed.

There are established techniques to measure the transverse polarization
of the positron based on the azimuthal distribution of
 the photon pair produced by the annihilation of the positron on the polarized
electrons in a thin magnetized target \cite{Burkard:1985wn}. We
have shown that a similar setup can be employed to measure the
transverse polarization of the electrons, too. In fact, the
azimuthal distribution of the final electrons in M\"{o}ller
scattering, $e_1^- e_2^- \to e_3^- e_4^-$ with polarized $e_2^-$
is sensitive to the polarization of $e_1^-$. The challenges before
this measurement are similar to the ones in \cite{Burkard:1985wn}
and can be overcome by similar methods.

  One can repeat similar discussion for three body LFV decays of
  $\tau$ lepton such as $\tau \to \mu \mu \mu$ or $\tau \to eee$.
  The measurements of the angular distribution and polarization of
  the final particles in these
decays can provide complementary information on the
$\lambda_{ijk}$ couplings. Such a study will be presented
elsewhere.

\section*{Acknowledgement}
S. N. would like to thank Prof. F. Arash for useful discussions.
She is also grateful to the physics department of IPM for the
hospitality of its staff and the partial financial support.

\end{document}